\documentclass[preprint,12pt]{elsarticle}
\journal{Nuclear Physics A}

\usepackage{graphicx}
\usepackage{epstopdf}
\usepackage{amssymb,amsmath,amsfonts}
\usepackage{afterpage}

\begin{document}

\begin{frontmatter}

%Title of paper
\title{Thermal, chemical and spectral equilibration in heavy-ion collisions}

\author{G\'abor Andr\'as Alm\'asi}
\ead{g.almasi@gsi.de}
\address[GSI]{Gesellschaft f\"{u}r Schwerionenforschung, GSI, D-64291 Darmstadt, Germany}
%\email[]{g.almasi@gsi.de}
\author{Gy\"{o}rgy Wolf}
\address[Wigner]{Wigner RCP, Budapest, Hungary}
\ead{wolf.gyorgy@wigner.mta.hu}
%\email[]{wolf.gyorgy@wigner.mta.hu}

\begin{abstract}
We have considered the equilibration in relativistic heavy ion collisions at energies 1-7 AGeV using our transport model. We applied periodic boundary conditions to close the system in a box. We found that the thermal equilibration takes place in the first 20-40 fm/c which time is comparable to the duration of a heavy ion collision. The chemical equilibration is a much slower process and the system does not equilibrate in a heavy ion collision. We have shown that in the testparticle simulation of the Boltzmann equation the mass spectra of broad resonances follow instantaneously their in-medium spectral functions as expected from the Markovian approximation to the Kadanoff-Baym equations employed via the (local) gradient expansion.
\end{abstract}


\end{frontmatter}

\section{Introduction \label{Introduction}}

In the past half century a great amount of effort was put into understanding the dynamics of relativistic heavy ion collisions. In these collisions only the initial and the final state is known experimentally, so the link between them needs to be provided by theoretical tools that describe the hot and dense medium created by the collision. The theoretical models are developed in two directions: In the thermal and hydrodynamical models \cite{BraunMunzinger1,Cleymans,Rischke,Rischke-dis,Heinz} some kind of equilibration is assumed, while in transport models \cite{Bertsch1,WolfBUUbasic0,GiBUU,Brat-Cassing,URQMD} the propagation of all particles is followed. In the first case some very simple, but rather strong assumptions are needed (global or local equilibration, thermal and chemical freeze out at equilibrium), and in the second case there are a lot of small details which should be approximated, and the model building and the calculations are rather tedious. It is desirable to compare the two approximations. For that purpose we use a transport-model calculation for a system, that is restricted to limited volume, which we will call the box in the following discussion. We study the timescales needed for thermal and for chemical equilibration of the box.

Another important question addressed in this paper is the modification of the properties of resonances in the nuclear medium. We focus on the vector mesons $\rho$ and $\omega$, since their in-medium properties are closely related to the experimentally observable dilepton spectrum. We expect their spectral function to undergo a significant change in the medium, hence we reckon the mass distribution of the created mesons to differ significantly from the vacuum spectral functions. How the mass distribution of these resonances change during the collision is a non-trivial question. In the year 2000 Cassing, Juchem \cite{Cassing1,Cassing2,Cassing3} and independently Leupold \cite{Leupold-trans} derived a system of equations from the Kadanoff-Baym equation, which approximately describes the propagation of resonances in a strongly interacting medium. These equations were implemented in several transport codes \cite{Wolfvector1,Wolfvector2,Cassing2,Bratkovskaya,GiBUU}. The propagation of the spectral function of vector mesons was studied by Schenke and Greiner \cite{Schenke}, too, by directly solving the Kadanoff-Baym equations without approximations. Here the background was a thermal medium with changing parameters. We studied the modification of the vector meson spectral functions in our transport model (restricted to a box) by applying the equations of Cassing, Juchem and Leupold and compared our results to the quantum field theoretical results of Schenke and Greiner.

Our paper is organized as follows. Essential features of our transport model are outlined in Section \ref{BUU}. The details and the numerical results of our box-simulations are presented in Section \ref{Modelling}. In Section \ref{Vector mesons} we study analytically and numerically the evolution of mass distributions of the vector mesons in thermal medium. Discussion and summary can be found in Section \ref{Summary}.\newpage

\section{The BUU model \label{BUU}}
The Boltzmann--Uehling--Uhlenbeck (BUU) model \cite{Bertsch1,WolfBUUbasic0,GiBUU,Brat-Cassing} is a microscopic transport model, which is based on the Boltzmann-equation and includes besides collisions an electromagnetic and a strong mean-field interaction between the particles. The equations are solved by approximating distributions by testparticles and using the parallel ensemble method. The elementary particles of our version of the model \cite{Wolfvector1,Wolfvector2} are baryons and mesons: $N(938)$, $N(1440)$, $N(1520)$, $N(1535)$, $N(1650)$, $N(1675)$, $N(1680)$, $N(1700)$, $N(1710)$, $N(1720)$, $N(2000)$, $N(2080)$, $N(2190)$, $N(2220)$, $N(2250)$, $\Delta(1232)$, $\Delta(1600)$, $\Delta(1620)$, $\Delta(1700)$, $\Delta(1900)$, $\Delta(1905)$, $\Delta(1910)$, $\Delta(1920)$, $\Delta(1930)$, $\Delta(1950)$, $\Lambda(1116)$, $\Sigma(1193)$, $\pi$, $\eta$, $\rho$, $\sigma$, $\omega$, $K^+/K^0$. The data of the resonances are partly taken from the Particle Data Group summary book\cite{PDG}, and partly from fits to $\pi N$ collision data. The $\sigma$ is a scalar-isoscalar meson, which simulates correlated pion pairs.

In the initial state of the program there are two heavy ions flying towards each other, and the final state of the program contains the particles leaving the collision. The evolution of the system is carried out in discrete time steps, until the interaction between the particles becomes negligible. In each time step we update the momentum and the coordinates of each particle, and test if there are any collisions. Two particles collide, if their distance in their center-of-momentum frame is smaller than $\sqrt{\frac{\sigma}{\pi}}$, where $\sigma$ is the total cross section of the collision. In the case of inelastic collision the particles can cease to exist and new particles can emerge from the center of their collision (while respecting the conservation laws). If in a reaction a resonance is produced together with another particle, then the energy-momentum conservation does not determine alone the mass of the resonance. The resonance mass is randomly chosen according to its spectral function and the available phase-space:
\begin{equation}
S(M_{res}) = \frac{\mathcal{A}(M_{res}) p_f(M_{res})}{\int dM_{res} \mathcal{A}(M_{res}) p_f(M_{res})},
\end{equation}
where $\mathcal{A}(M_{res})$ is the spectral function of the resonance and $p_f$ is the outgoing resonance momentum in the center of mass frame of the microscopical collision.

 The interactions incorporated in the program are shown in Table \ref{partcoll}.
\begin{table}[h]
\centering
%\begin{ruledtabular}
\begin{tabular}{ | c | c | }
  \hline
  $NN \leftrightarrow NN$ & $NR \leftrightarrow NR$  \\ \hline
  $NN \leftrightarrow NR$ & $NR \leftrightarrow NR'$  \\ \hline
  $NN \leftrightarrow \Delta(1232)\Delta(1232)$ & $R \leftrightarrow N m$  \\ \hline
  $R \leftrightarrow \Sigma K$ & $R \leftrightarrow \Lambda K$  \\ \hline
  $R \leftrightarrow \Delta(1232)\pi$ &  $R \leftrightarrow N(1440)\pi$ \\ \hline
  $\rho \leftrightarrow \pi\pi$ & $\sigma \leftrightarrow \pi\pi$ \\ \hline
  $\omega \leftrightarrow \rho\pi$ & $NN \leftrightarrow NN\pi$  \\ \hline
\end{tabular}
%\end{ruledtabular}
\caption{\label{partcoll} The types of collisions implemented in the model. The $N$ without mass specification means nucleons, $R$ and $R'$ denote arbitrary baryon resonances, while $m$ denotes an arbitrary nonstrange meson ($\pi, \eta, \sigma, \rho, \omega$).}
\end{table}

Between collisions the testparticles move in an electromagnetic and a nuclear mean-field potential. The electromagnetic potential is the sum of the Coulomb potentials of the charged testparticles. It affects the motion of all charged particles. The nuclear mean-field potential only interacts with the baryons, and depends only on the local baryon-density and the momentum of the testparticle. The concrete form of the interaction is the following:
\begin{equation}
	U^{nr}=A\frac{\rho}{\rho_0} + B \left(\frac{\rho}{\rho^0}\right)^{\tau} 
	+ \frac{2C}{\rho_0}\int \frac{d^3\mathbf{p}'}{(2\pi^3)} \, 
	\frac{f(\mathbf{r},\mathbf{p}')}{1+\left(\frac{\mathbf{p}-\mathbf{p}'}{\Lambda}\right)^2},
\end{equation}
where $\rho$ is the local baryon density, $\rho_0=0.168\; 1/fm^3$ is the normal nuclear density, $f(\mathbf{r},\mathbf{p})$ is the distribution function of nucleons. The parameters of the model are given by: $A=-26.09\, \textrm{MeV}$, $B=56.59\, \textrm{MeV}$, $C=-64.65\, \textrm{MeV}$, $\tau=1.764$ and $\Lambda = 2.168\, \textrm{fm}^{-1}$. These parameters are obtained by a fit to the binding energy, to normal nuclear density, to compressibility and to the nuclear optical potential properties \cite{WolfTeis}. The local baryon density is calculated in the following way: we assign a Gaussian density distribution with $1\, \textrm{fm}$ width to all testparticles, and we calculate the total density coming from these contributions on a $1\, \textrm{fm}$ cubic grid. The program calculates the electric charge density the same way.
This potential is the generalization of the momentum independent Skyrme-potential. The momentum dependent part of the interaction is basically a Yukawa coupling with a scalar particle with mass $\Lambda$. Since the potential used is clearly nonrelativistic, problems can emerge from the fact that we carry out the propagation of particles in laboratory frame, and calculate collisions in center-of-momentum frame of the colliding particles. Instead of using the nonrelativistic potential, we define the following potential $U$ in the center-of-momentum frame of the matter flowing around the testparticle:
\begin{equation}
	\sqrt{\mathbf{p}^2+m^2} + U^{nr}(\mathbf{r},\mathbf{p}) = \sqrt{\mathbf{p}^2 + (m + U(\mathbf{r},\mathbf{p}))^2}.
\end{equation}
We treat $U$ as if it were a relativistic scalar potential, and this way we can calculate the potential in any frame.
The propagation of the testparticles is carried out with predictor-corrector method. More details of the model can be found in \cite{Wolfvector1}.

\section{\label{Modelling} Modeling of strongly interacting matter in equilibrium}

By introducing periodic boundary  conditions for outgoing particles of a heavy ion collisions, we can model strongly interacting matter on hadronic level. We initialize the system from a ${}^{12}C+{}^{12}C$ collision. After the collision take place, we introduce periodic boundary conditions, so that particles coming from the center of the collision cannot leave a finite volume. For simplicity we choose this finite volume to have cubic shape, and from now on we will refer to it as the box.

It is important to investigate if the created system gives a good model of strongly interacting matter. Since the equations of the BUU model conserve baryon number, it cannot change over time. However nothing guarantees the conservation of energy, or the number of mesons. The equations of the BUU model do conserve energy, (although not with the ``off-shell propagation'' explained later) but numeric approximating methods that solve these equations do not have this pleasant property. In our model we demand that our system has constant total energy over time. If our system fulfills this demand, we would like to observe the process of thermalization and chemical equilibration.

\subsection{The periodic boundary condition}

In our model we employ periodic boundary conditions in order to keep the outgoing particles in the box. We contain the particles in the following volume:
\begin{eqnarray}
	-a/2 &< x \leq a/2, \nonumber \\
	-a/2 &< y \leq a/2,  \\
	-a/2 &< z \leq a/2, \nonumber
\end{eqnarray}
where $a$ denotes the size of the box. We introduce this boundary condition in the following way: if a particle gets outside of the box during propagation, we shift the appropriate coordinate of the particle by $a$, so that the particle gets back to (the other side of) the box. Instead of changing the coordinate, we could have mirrored the momentum of the particle  to the wall of the box, imitating an elastic collision with the wall, but because of the momentum dependent potential, this would break the energy conservation.

In Section \ref{BUU} we mentioned, when calculating the values of the mean-field potentials, the program evaluates baryon and electric charge densities on a $1\, \textrm{fm}$ cubic grid. When a particle leaves the box and re-enters it on its other side, we would like its momentum to change smoothly, and the total energy to be conserved. This means that the mean-field potentials have to change continuously in these transitions, and the density functions have to obey the following relations:
\begin{eqnarray}
	\rho(a/2,y,z)=\rho(-a/2,y,z)\quad \forall y,z, \nonumber \\
	\rho(x,a/2,z)=\rho(x,-a/2,z)\quad \forall x,z, \\
	\rho(x,y,a/2)=\rho(x,y,-a/2)\quad \forall x,y. \nonumber
\end{eqnarray}
These equations can be easily fulfilled by calculating densities at all points of the grid (also outside of the box). Then in a given point we carry out the following summation:
\begin{equation}
	\rho '(x,y,z)=\sum_{i=-1}^1\sum_{j=-1}^1\sum_{k=-1}^1 \rho(x+i\cdot a,y+j\cdot a,z+k\cdot a).
\end{equation} 
This way the densities will be equal at arbitrary opposing sides of the box, and there will not be any discontinuities during propagation of particles. The summation of densities can be justified the following way: in an infinite medium particles outside of the box would interact with the particles inside the box and if represented by Gaussian density profile, the exterior particles would modify the density inside the box. By using periodic boundary conditions the opposing side of the box acts as matter outside of the box. During our work we used a cubic box with edge length $a=5\, \textrm{fm/c}$. When choosing edge length one should remember that the densities are calculated on a $1\, \textrm{fm}$ grid, and periodic boundary conditions demand that a grid point shifted by $a$ parallel to any grid axis should fall on another grid point.

\subsection{Stability of the system}

Having defined the initial condition of the system and the boundary conditions, now we turn to the stability of the system we created. Since we would like to give a model of strongly interacting matter, we wish to keep the temperature of the system and the particle content roughly constant. Hopefully this can be achieved by fixing the total energy of the system. The total energy of the system can be divided into five distinct parts: kinetic energy of mesons, kinetic energy of baryons, rest mass of mesons, rest mass of baryons and the energy of the fields. Since the mean-fields of the BUU model can change the mass of particles, the definition of the previous energies are not trivial. We chose to define the energies the following way:
\begin{align}
\begin{split}
	E_{barkin}&=\sum_i \sqrt{\mathbf{p}_i^2 + \left( m_i + U_i\right)^2} - (m_i + U_i),  \\
	E_{barrest}&=\sum_i (m_i - m_{nuc}),  \\
	E_{meskin}&=\sum_i \sqrt{\mathbf{p}_i^2 + m_i^2} - m_i,  \\ 
	E_{mesrest}&=\sum_i m_i,
\end{split}
\end{align}
where the $i$ index runs over all particles, $m_i$ denotes the rest mass of the given particle in vacuum, $m_{nuc}$ denotes the rest mass of nucleons in vacuum, $\mathbf{p}_i$ is the momentum of the particle and $U_i$ is the value of the nuclear mean-field potential at the location of the particle, if the particle is a baryon. We defined the rest energy of the baryons this way, because the total number of baryons is fixed, so we removed a constant shift by subtracting $m_{nuc}$ from every baryon mass. In this total rest energy only baryon resonances contribute.

%\begin{minipage}{\linewidth}
\begin{figure}[h]
   \includegraphics[width=0.8\linewidth]{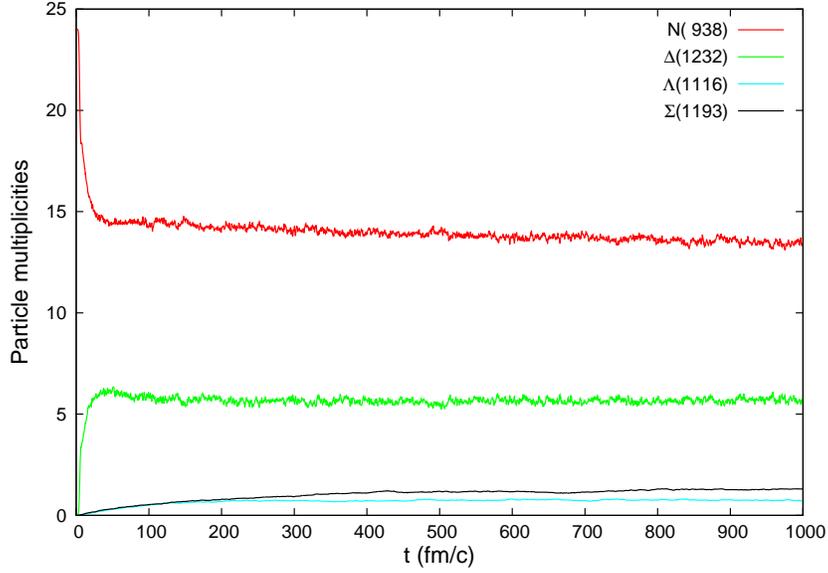}
   \caption{ Number of baryon resonances plotted against time in a run initialized from a $3\, \textrm{AGeV}$ $C+C$ collision. \label{baryons_nofixe}}
\end{figure}
\begin{figure}[h]
   \includegraphics[width=0.8\linewidth]{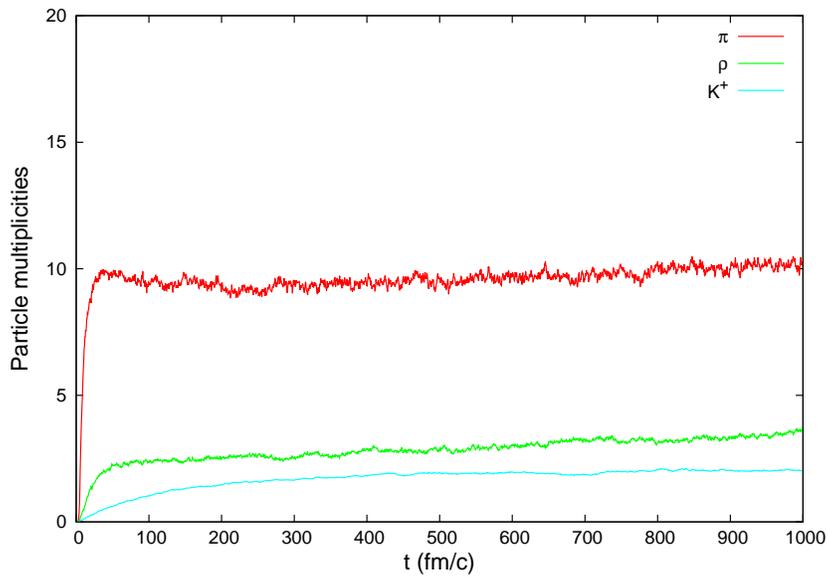}
   \caption{ Number of mesons plotted against time in a run initialized from a $3\, \textrm{AGeV}$ $C+C$ collision. \label{mesons_nofixe}}
\end{figure}
\begin{figure}[h]
   \includegraphics[width=0.8\linewidth]{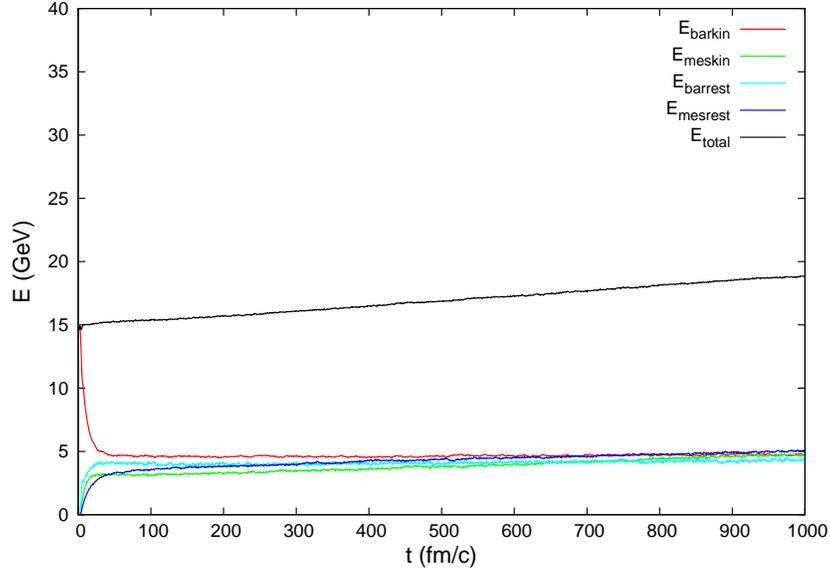}
   \caption{ Kinetic energy and rest mass of mesons and baryons plotted against time in a run initialized from a $3\, \textrm{AGeV}$ $C+C$ collision. \label{energies_nofixe}}
\end{figure}
\begin{figure}[h]
   \includegraphics[width=0.8\linewidth]{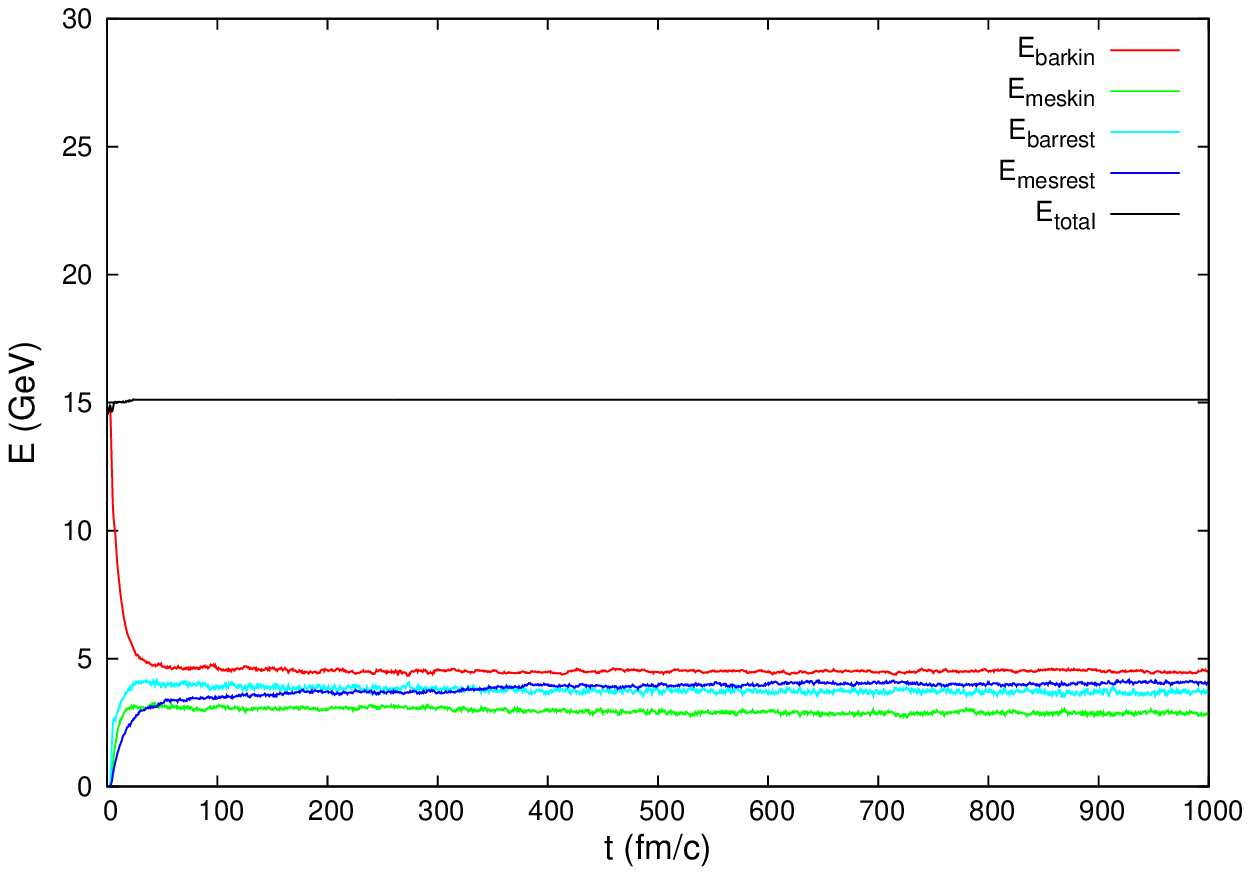}
   \caption{ Kinetic energy and rest mass of mesons and baryons plotted against time in a run initialized from a $3\, \textrm{AGeV}$ $C+C$ collision, with fixed total energy. \label{energies_fixe}}
\end{figure}
%\end{minipage}
\afterpage{\clearpage}

We plotted the above defined energies against time in a run initialized with a $3\, \textrm{AGeV}$ collision. In Fig. \ref{baryons_nofixe} and \ref{mesons_nofixe} we can see that under $1000\, \textrm{fm/c}$ the average value of particle numbers changes significantly, the system is not chemically stable. In Fig. \ref{energies_nofixe} we can see that the total energy of the system is not constant, it changes significantly under $1000\, \textrm{fm/c}$. Note however, that the change of the total energy in real collision processes which take not more than $40\, \textrm{fm/c}$ is negligible. There are two sources of this effect. Equation (\ref{Eq-Hamilton}), which we are solving is only approximately energy conserving \cite{Leupold-trans,Knoll}, and the numerical approximations of our model using momentum dependent interaction also lead to small changes in the energy. This is an effect we want to get rid of when performing very long simulations. The easiest way to do it is to correct the total energy of the system in each time step to a fixed value by hand. It is reasonable to apply modifications only on baryons in order to fix the total energy. Denoting the loss of total energy per baryon by $\Delta E$ in a time step, we transform the momentum of all baryons the following way to fix the total energy:
\begin{equation}
	\mathbf{p}_i \rightarrow \mathbf{p}'_i =
	\frac{\sqrt{\left(\Delta E +\sqrt{\mathbf{p}_i^2+(m_i+U_i)^2}\right)^2-\left(m_i+U_i\right)^2}}
	{\left|\mathbf{p}_i\right|}
	\cdot\mathbf{p}_i,
\end{equation}
since this way all baryon particles gain $\Delta E$ energy:
\begin{eqnarray}
	E'_i=\sqrt{\mathbf{p}_i'^{\, 2}+\left(m_i+U_i\right)^2} &=& \sqrt{\mathbf{p}_i^2+(m_i+U_i)^2} + \Delta E \nonumber \\ &=&
	E_i + \Delta E.
\end{eqnarray}
This modification cannot be done in all cases, since $\Delta E$ can be negative, so the factor multiplying the momenta can be imaginary. If the modification cannot be done, then we add the $\Delta E$ energy to another, randomly chosen baryon.

\subsection{Thermalization of the system}

By fixing the energy of the system with the previously described method we get a stable system. Our aim is to approximately determine the time needed for thermalization. For this purpose we investigate first how quickly the ratios of different energy types and the particle content become constant. We plotted the above defined energies and particle content against time in a run initialized from a ${}^{12}C+{}^{12}C$ collision now with fixed total energy. Using this initialization the average density in the box will be $0.192\, \textrm{fm}^{-3}$ which is somewhat larger than the normal nuclear density.

\begin{figure}[t]
\begin{minipage}[t]{0.48 \linewidth}
   \includegraphics[width=\linewidth]{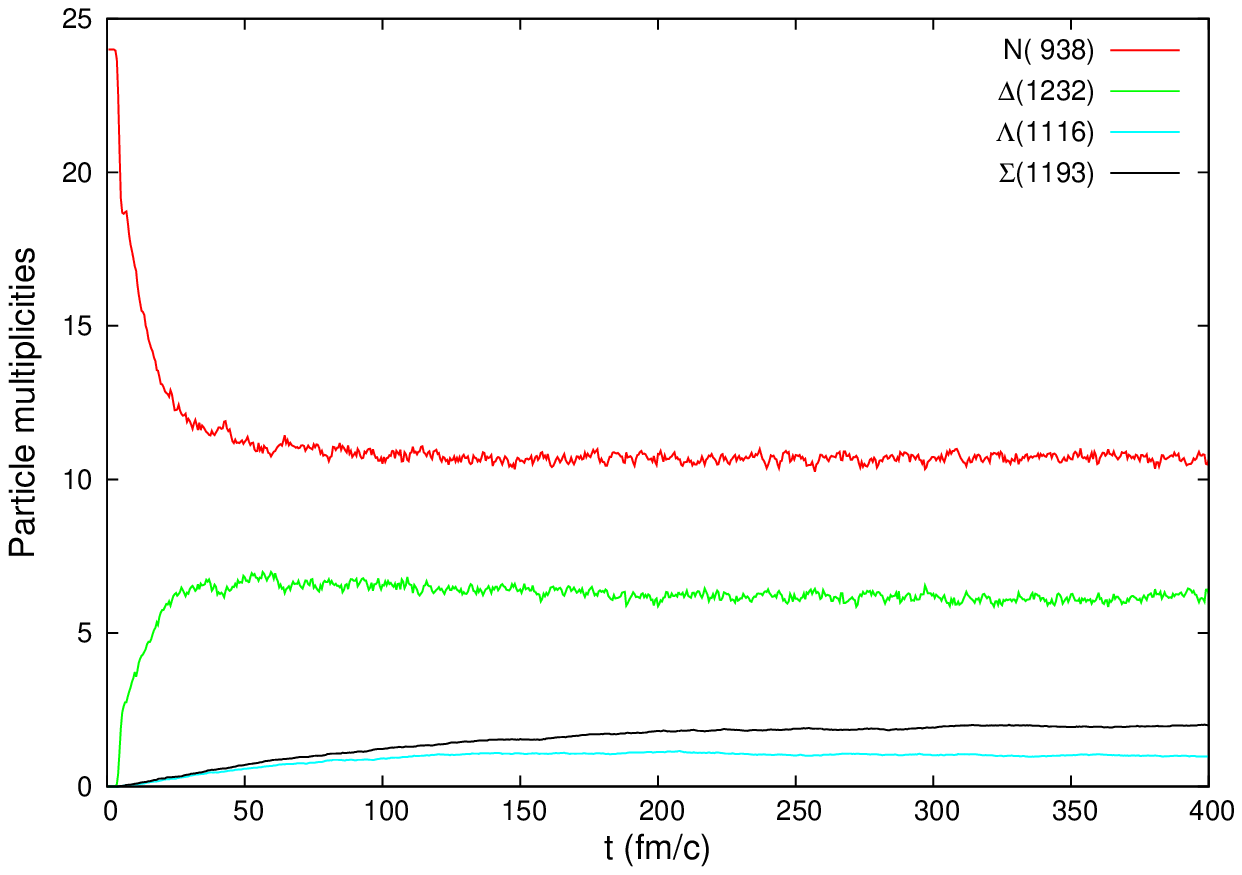}
   \caption{ Number of baryon resonances plotted against time in a run initialized from a $7\, \textrm{AGeV}$ $C+C$ collision, with fixed total energy. \label{baryons_fixe}}
\end{minipage}
\hspace{1cm}
\begin{minipage}[t]{0.48 \linewidth}
   \includegraphics[width=\linewidth]{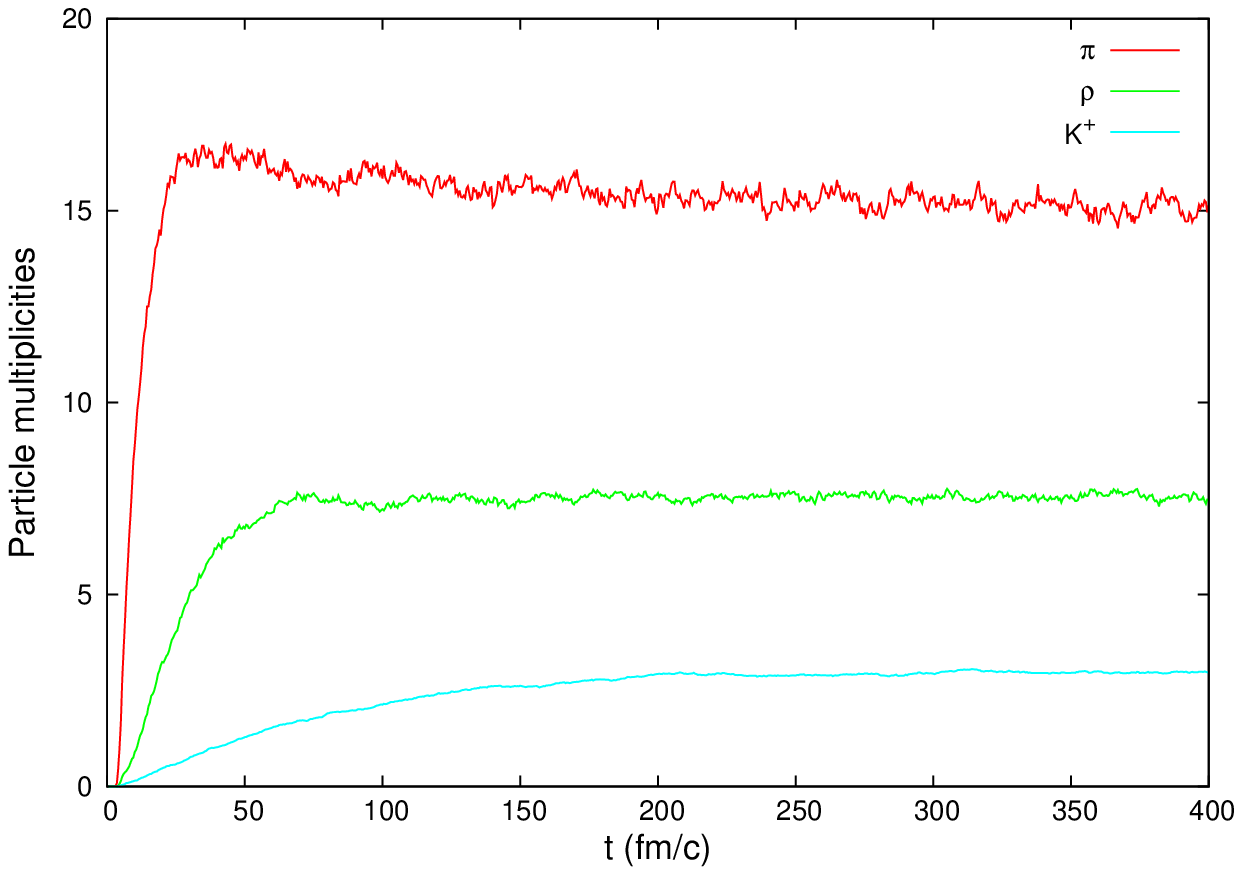}
   \caption{ Number of mesons plotted against time in a run initialized from a $7\, \textrm{AGeV}$ $C+C$ collision, with fixed total energy. \label{mesons_fixe}}
\end{minipage}
\begin{minipage}[t]{0.48 \linewidth}
      \includegraphics[width=\linewidth]{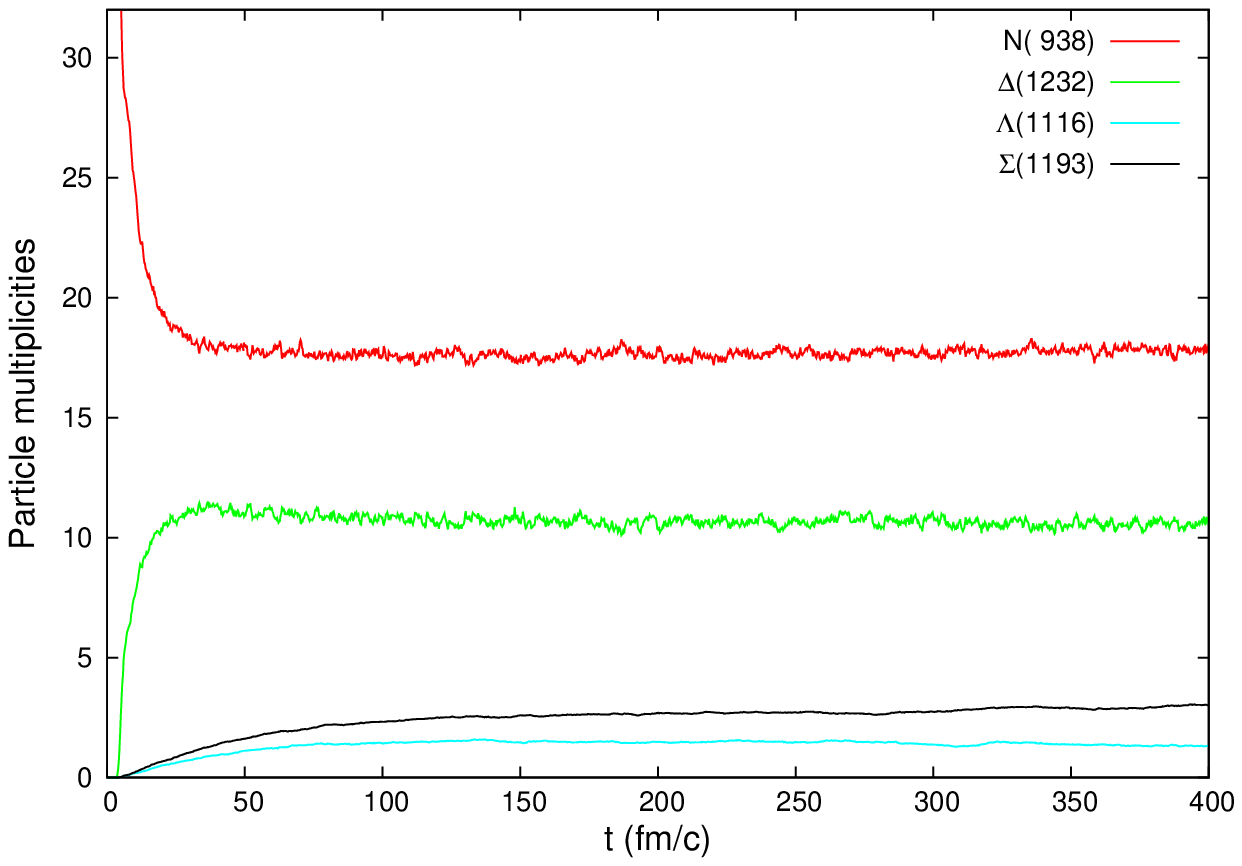}
   \caption{ Number of baryon resonances plotted against time in a run initialized from a $7\, \textrm{AGeV}$ $Ca+Ca$ collision, with fixed total energy. \label{baryons_CaCa}}
\end{minipage}
\hspace{1cm}
\begin{minipage}[t]{0.48 \linewidth}
   \includegraphics[width=\linewidth]{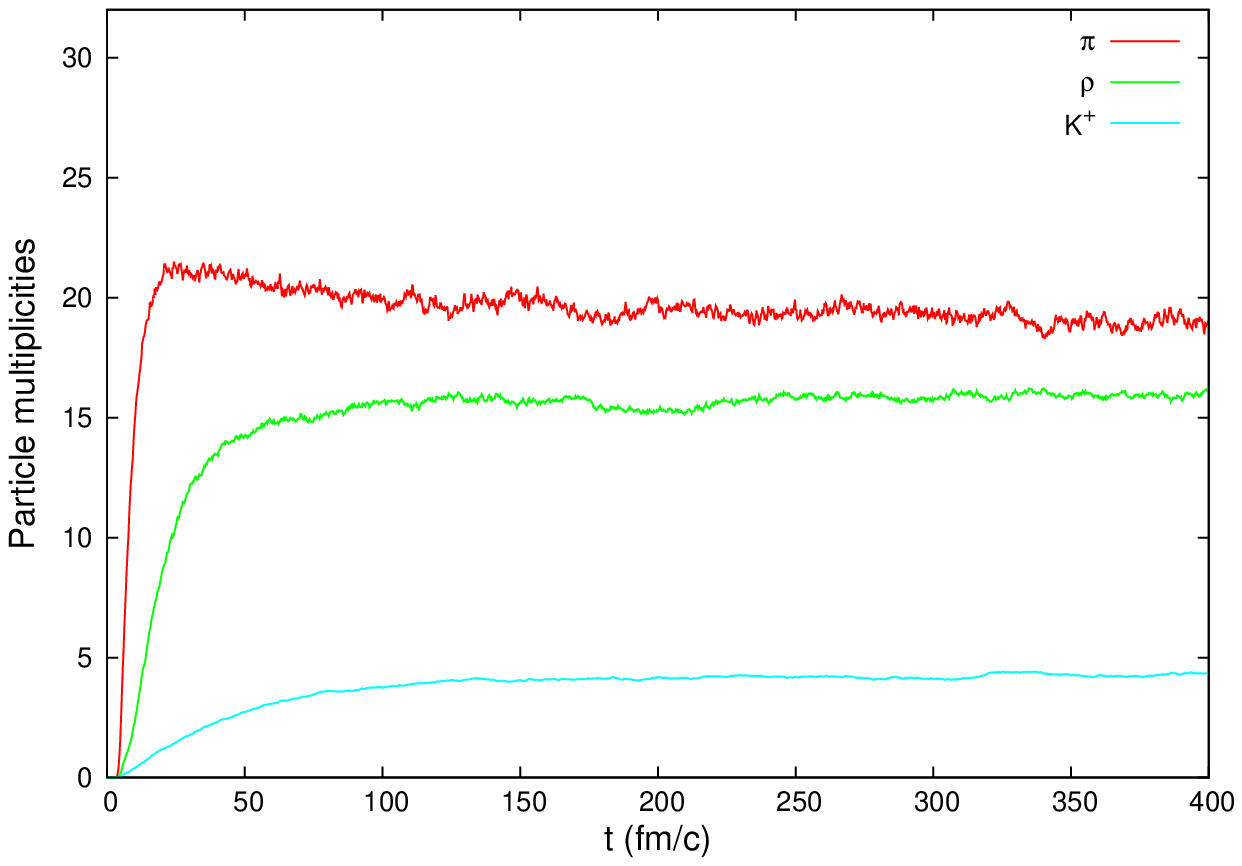}
   \caption{ Number of mesons plotted against time in a run initialized from a $7\, \textrm{AGeV}$ $Ca+Ca$ collision, with fixed total energy. \label{mesons_CaCa}}
\end{minipage}
\end{figure}

In Fig. \ref{energies_fixe} we can see that in a $3\, \textrm{AGeV}$ collision the ratios of the average value of different energy types become constant in about $200\, \textrm{fm/c}$, and after this transition time the ratios of the energies fluctuate around a constant value. In Figs. \ref{baryons_fixe} and \ref{mesons_fixe} we can see that in a run initialized with a $7\, \textrm{AGeV}$ collision the number of particles with strangeness stabilizes also in about $200\, \textrm{fm/c}$. The number of particles without strangeness seems to get approximately constant much faster, but the conservation laws of energy and baryon number relate the nonstrange with the strange sector, so the final thermalization of the nonstrange particles happens at the same time when the strange particles thermalize. When we decrease the energy by initializing the system from $3\, \textrm{AGeV}$ and $1\, \textrm{AGeV}$ ${}^{12}C+{}^{12}C$ collisions, we find that the equilibration time increases approximately to $300\, \textrm{fm/c}$ for the $3\, \textrm{AGeV}$ and to $500\, \textrm{fm/c}$ for the $1\, \textrm{AGeV}$ collisions.

We also investigate what happens if we change the average density by initializing the system from a ${}^{40}Ca+{}^{40}Ca$ collision instead of a ${}^{12}C+{}^{12}C$ collision with the same $7\, \textrm{AGeV}$ colliding energy. In this case the average density of the system is $0.64\, \textrm{fm}^{-3}$ which is around four times the normal nuclear density. At this density the chemical equilibration time reduces to $100\, \textrm{fm/c}$
(see Figs. \ref{baryons_CaCa} and \ref{mesons_CaCa}), which is still much larger than the whole reaction time in a real heavy ion collision at this energy.

\begin{figure}[h!]
\begin{minipage}[t]{0.48 \linewidth}
   \includegraphics[width=\linewidth]{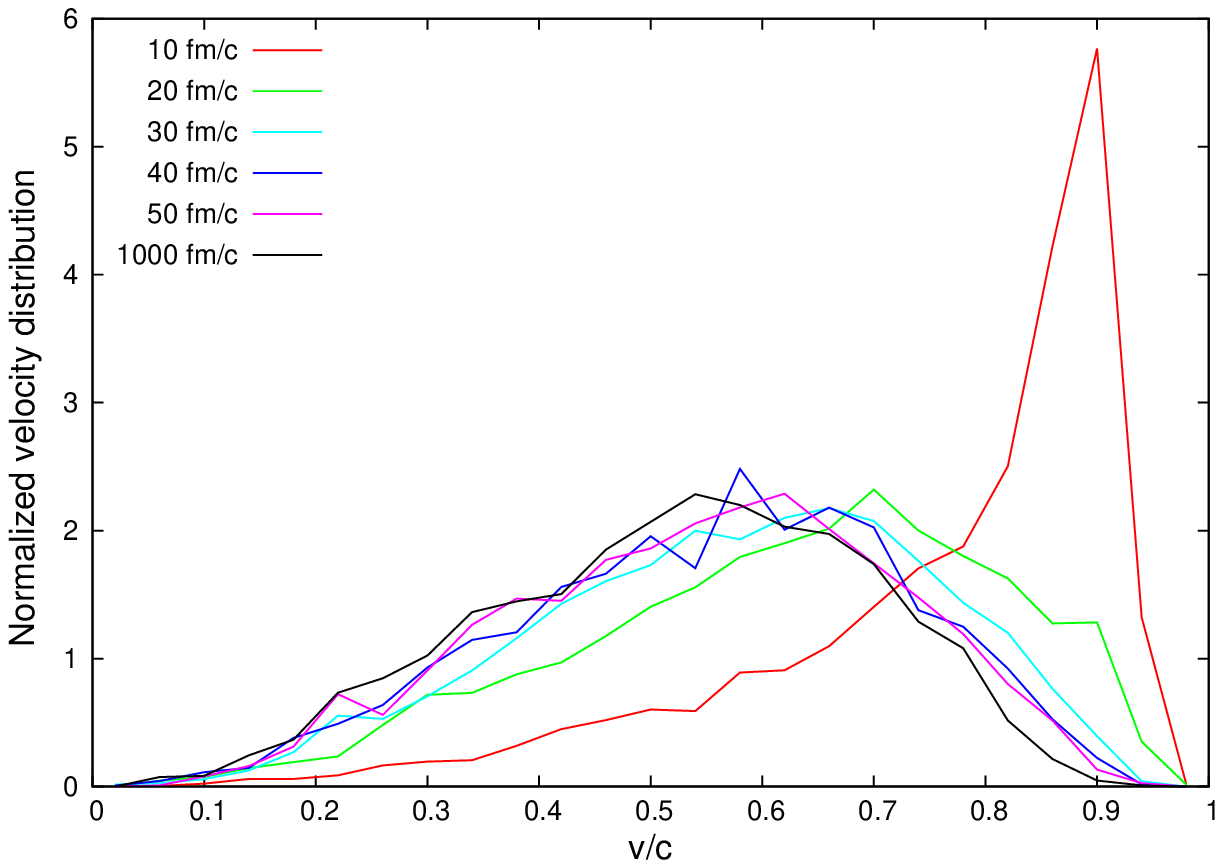}
   \caption{The velocity distribution of nucleons at different times in a run initialized from a $7\, \textrm{AGeV}$ $C+C$ collision.}  \label{nuklvelodist}
\end{minipage}
\hspace{1cm}
\begin{minipage}[t]{0.48 \linewidth}
   \includegraphics[width=\linewidth]{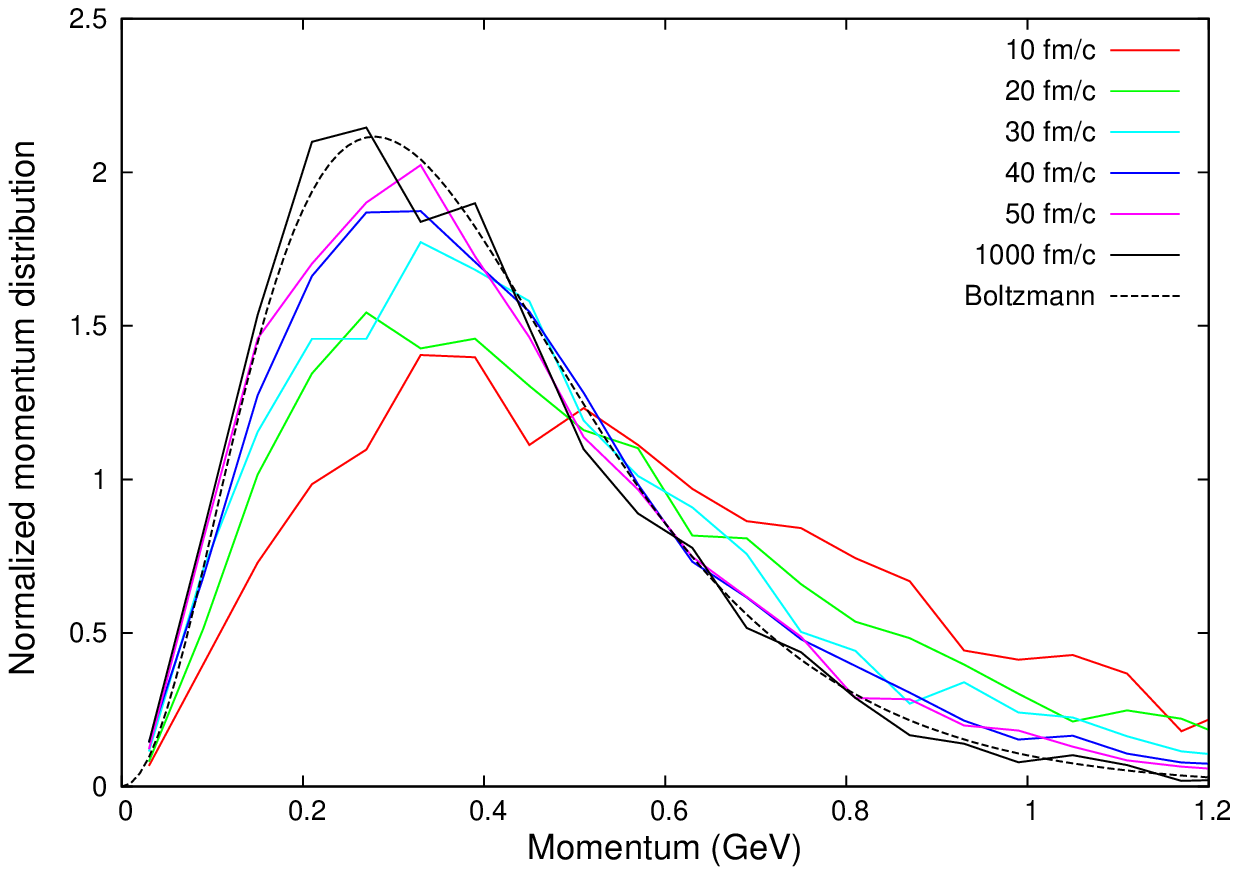}
   \caption{The momentum distribution of pions at different times in a run initialized from a $7\, \textrm{AGeV}$ $C+C$ collision. The dashed line show the Boltzmann distribution with T=125 MeV.} 
   \label{pionmomdist}
\end{minipage}
\begin{minipage}[t]{0.48 \linewidth}
   \includegraphics[width=\linewidth]{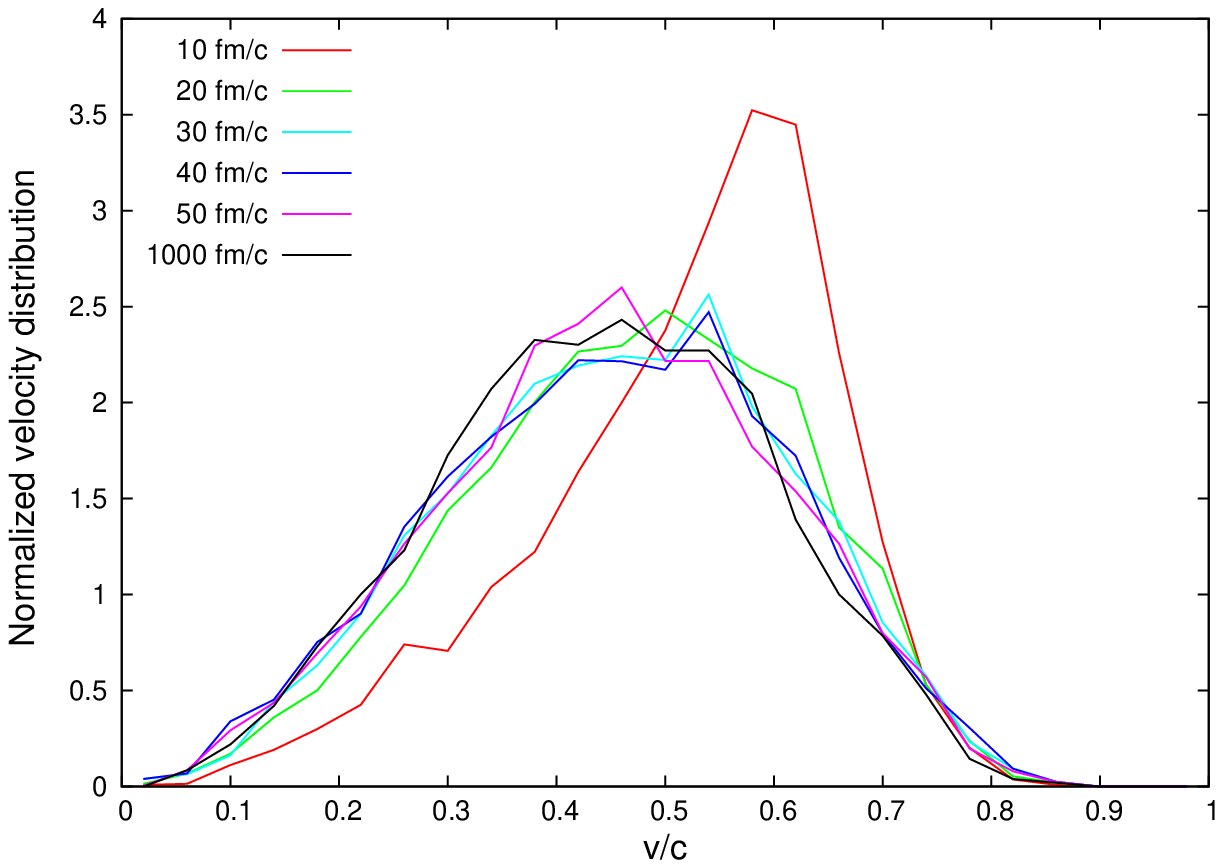}
   \caption{The velocity distribution of nucleons at different times in a run initialized from a $1\, \textrm{AGeV}$ $C+C$ collision.} 
   \label{nuklvelodist_fixeE1}
\end{minipage}
\hspace{1cm}
\begin{minipage}[t]{0.48 \linewidth}
   \includegraphics[width=\linewidth]{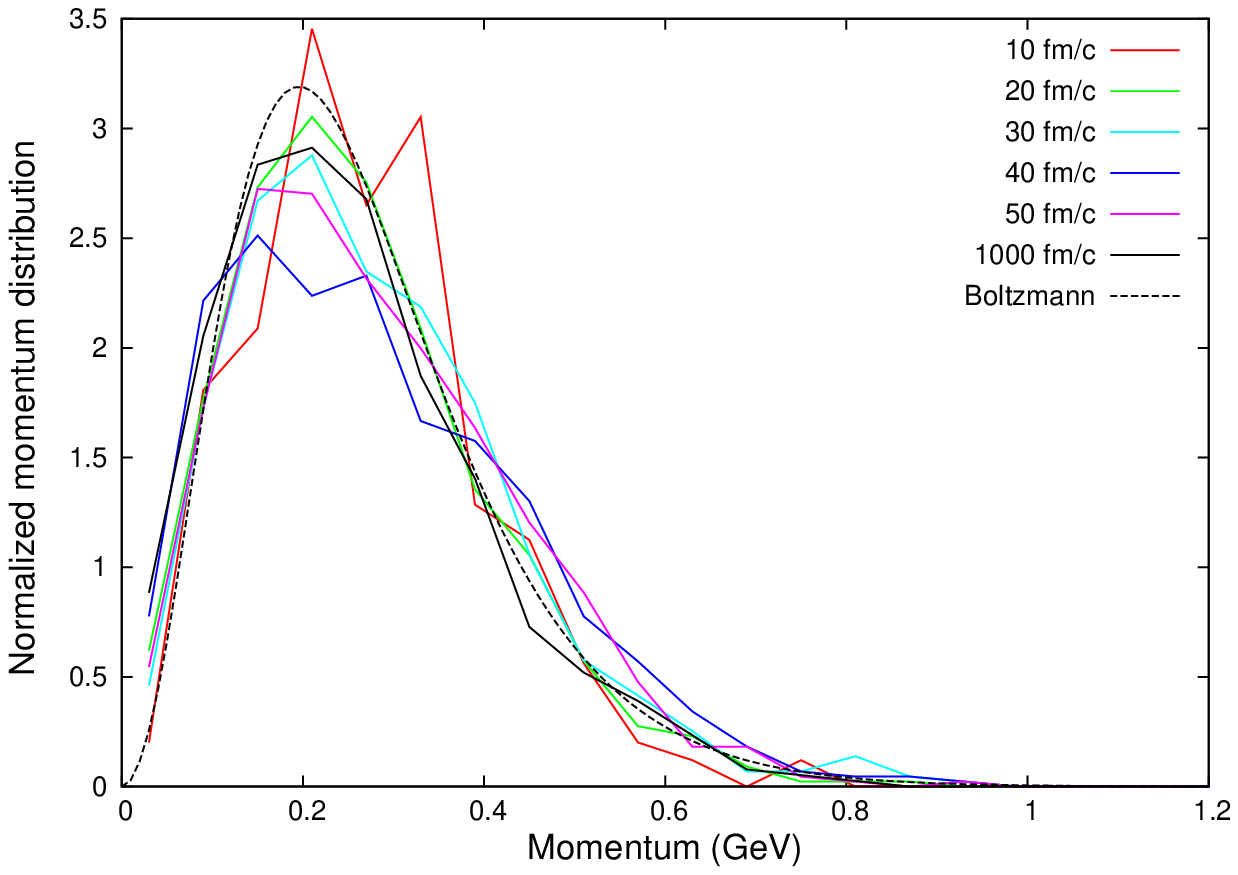}
   \caption{The momentum distribution of pions at different times in a run initialized from a $1\, \textrm{AGeV}$ $C+C$ collision. The dashed line show the Boltzmann distribution with T=80 MeV.} 
   \label{pionmomdist_fixeE1}
\end{minipage}
\end{figure}

\begin{figure}[t] 
 \centering
   \includegraphics[width=\linewidth]{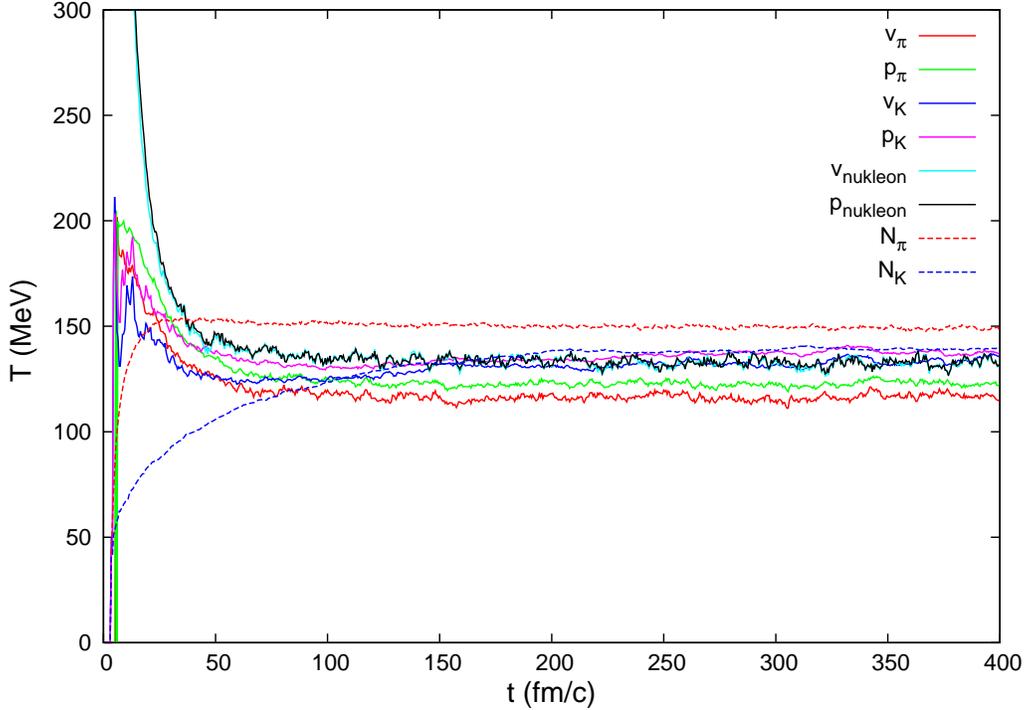}
   \caption{The temperature of the system extracted by particle contents and momentum distributions initialized from a $7\, \textrm{AGeV}$ $C+C$ collision.} 
   \label{combinedtemp}
\end{figure}

Thus we can see that the full chemical equilibration takes more time, than the average time of a heavy ion collision. Kinetic equilibration can however be much faster. To examine this we plotted the velocity distribution of nucleons and the momentum distribution of pions at the initial stage of the thermalization in a system initialized from a $7\, \textrm{AGeV}$ $C+C$ collision. In Fig. \ref{nuklvelodist} we see that the velocity distribution of the nucleons becomes constant after around $30 \text{--} 40\, \textrm{fm/c}$, which is comparable to the timescale of heavy ion collisions. In Fig. \ref{pionmomdist} we see that the same is true for the momentum distribution of pions. When increasing the density by initializing from a $7\, \textrm{AGeV}$ $Ca+Ca$ collision we found no substantial change in the kinetic equilibration time. On the other hand, if we decrease the initial collision energy to $1$ or $3\, \textrm{AGeV}$, the kinetic equilibration time reduces to approximately $20 \textrm{fm/c}$ as one can see in Figs. \ref{nuklvelodist_fixeE1} and \ref{pionmomdist_fixeE1}. In Figs. \ref{pionmomdist} and \ref{pionmomdist_fixeE1} we show that the pion momentum distributions converge to a Boltzmann-distribution with T=80 MeV and T=125 MeV, respectively.
We can determine the temperature of the system from the particle content or from the velocity or momentum distributions by fitting to the corresponding quantities of the free hadron resonance gas. Since our system is not a free gas, different methods of temperature calculations give different results.  We find that this uncertainty in the temperature is less than 15\% (Fig. \ref{combinedtemp}). The temperature of the pions falls somewhat out, but the process $NN\leftrightarrow NN\pi$ may not fulfill the detailed balance completely, and this explains that the pion number and momentum distribution give somewhat different temperatures than the rest.

 Similar investigation have been carried out in \cite{BelkacemEOS,Bravina2000,BratkovskayaEquilibration}. In \cite{BelkacemEOS,Bravina2000} they studied thermal and chemical equilibrations at higher energies relevant for AGS and SPS compared to our studies which is appropriate for SIS. In \cite{BratkovskayaEquilibration} the low energy case is studied as well, however, the correspondence between the thermal and chemical equilibration temperatures are not investigated like in Fig. \ref{combinedtemp} in this paper.

\subsection{Simulation of varying density nuclear matter}

\begin{figure}[p] 
 \centering
   \includegraphics[width=0.8\linewidth]{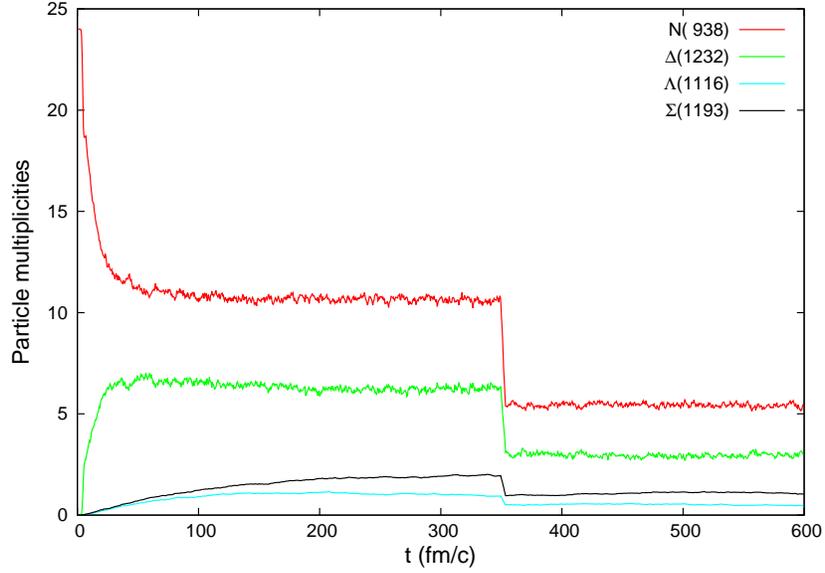}
   \caption{The number of baryon testparticles as a function of time. The density was decreased linearly by a factor of two under $\Delta t=4\, \textrm{fm/c} $ after $350\, \textrm{fm/c}$ of thermalization.}
   \label{baryons_chdens}
\end{figure}

\begin{figure}[p] 
 \centering
   \includegraphics[width=0.8\linewidth]{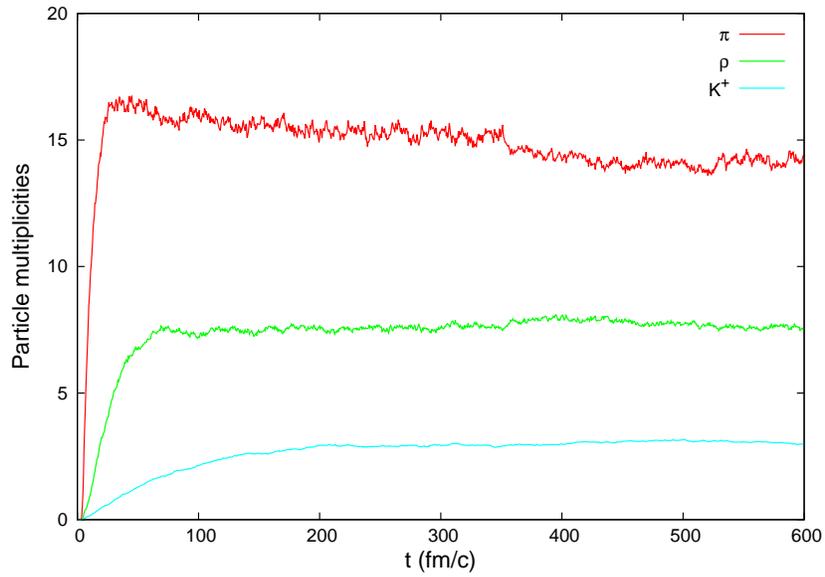}
   \caption{The number of meson testparticles in the function of time. The density was decreased linearly by a factor of two under $\Delta t=4\, \textrm{fm/c} $ after $350\, \textrm{fm/c}$ of thermalization.}
   \label{mesons_chdens}
\end{figure}

In heavy ion collisions the density and the temperature of the nuclear matter change rapidly. The rapidly changing density can modify the spectral functions of vector mesons, which motivates us to try to simulate nuclear matter with non-constant density. In the present work we aim to modify density linearly in time, leaving the temperature of the system approximately unchanged. The modification of density in our model can be accomplished by removing baryons from the system. In Fig. \ref{baryons_chdens} and \ref{mesons_chdens} we can see how the particle multiplicities change when the density is reduced linearly by a factor of two under $\Delta t=4\, \textrm{fm/c} $. After the reduction of density no transient effects seem to happen, the multiplicities do not change considerably. The number of $\rho$ mesons and pions however seem to change a little, which is caused by the change of the self-energy of the $\rho$ mesons.
\afterpage{\clearpage}
% \begin{widetext}
\section{\label{Vector mesons} Resonances in dense matter}

An interesting application of our nuclear medium is the investigation of its effect on the properties of broad resonances.  Similar numerical investigation was carried out in \cite{Cassing3}; however the equilibration of the mass distribution was not studied there, and this is our aim here. As an example we study the vector mesons, however, it applies similarly for other resonances as well. The dilepton production of heavy ion collisions depends strongly on the spectral functions of the vector mesons, therefore it is important to have a good understanding of the modification of their spectral functions in medium. We investigate this question by putting high number of vector meson testparticles ($\rho$ and $\omega$) in our box of interacting hadrons. We forbid their decay and any collisions between them: they only interact with the medium via their density dependent self-energies.

For the vector meson self-energies we chose the following form:
\begin{align}
\begin{split}
\label{selfen}
	\mathrm{Re}\Sigma(\mathbf{r},\mathbf{p},E,t) &= \mathrm{Re}\Sigma(\mathbf{r},t) = - (M_0^2-M_x^2) \frac{\rho(\mathbf{r},t)}{\rho_0}, \\
	\mathrm{Im}\Sigma(\mathbf{r},\mathbf{p},E,t) &= \mathrm{Im}\Sigma(\mathbf{r},t) =
	 - M_0\left((\Gamma_x-\Gamma_0) \frac{\rho(\mathbf{r},t)}{\rho_0} + \Gamma_0 \right),
\end{split}
\end{align}
where $M_0$ is the peak value of the mass distribution in vacuum, $\Gamma_0$ is the vacuum width, $\rho(\mathbf{r},t)$ is the local baryon density of the medium, $\rho_0 $ is the average density in the box at the initial state, whereas $M_x$ and $\Gamma_x$ are parameters to be chosen. The parameters $M_x$ and $M_0/M_x\:\Gamma_x$ correspond to the peak location and width of the mass distribution in a medium with a density $\rho(\mathbf{r},t) = \rho_0$. For the vacuum masses and widths we used the following values for the vector mesons: $M_0^{\rho}=768\, \textrm{MeV}$, $\Gamma_0^{\rho}=150\, \textrm{MeV}$, $M_0^{\omega}=782\, \textrm{MeV}$ and $\Gamma_0^{\omega}=8.5\, \textrm{MeV}$. Although in our model we use energy-dependent widths for $\rho$ and $\omega$ mesons, in this calculations we use energy independent width to avoid unnecessary complications caused by $C_i$ in Eq. (\ref{Eq-Hamilton}) which can be sometimes close to one in case of energy dependent width.

The masses of the vector meson testparticles are chosen randomly according to a Breit-Wigner distribution:
\begin{equation}
\label{spectfunc}
	A(M^2,\mathbf{r},t) = -\frac{2}{\pi}\frac{\mathrm{Im}\Sigma(\mathbf{r},t)}
	{\left(M^2-M_0^2-\mathrm{Re}\Sigma(\mathbf{r},t)\right)^2+\mathrm{Im}\Sigma(\mathbf{r},t)^2}.
\end{equation}
The time evolution of their coordinates are carried out according to the following equations \cite{Leupold-trans,Cassing3}:
\begin{align}
\label{Eq-Hamilton}
	\frac{d\mathbf{X}_i}{dt} &=
	 \frac{1}{1-C_i}\frac{1}{2 E_i}\left(2\mathbf{P}_i+\nabla_{\mathbf{P}_i}\mathrm{Re}\Sigma^{ret}_i 
	+ \frac{E_i^2-\mathbf{P}_i^2-M_0^2-\mathrm{Re}\Sigma^{ret}_i}{\mathrm{Im}\Sigma^{ret}_i}\,\nabla_{\mathbf{P}_i}\mathrm{Im}\Sigma^{ret}_i\right),
	\nonumber \\
	\frac{d\mathbf{P}_i}{dt} &= -\frac{1}{1-C_i}\frac{1}{2 E_i}\left(\nabla_{\mathbf{X}_i}\mathrm{Re}\Sigma^{ret}_i 
	+ \frac{E_i^2-\mathbf{P}_i^2-M_0^2-\mathrm{Re}\Sigma^{ret}_i}{\mathrm{Im}\Sigma^{ret}_i}\,\nabla_{\mathbf{X}_i}\mathrm{Im}\Sigma^{ret}_i\right),
	\\
	\frac{d E_i}{dt} &=
	 \frac{1}{1-C_i}\frac{1}{2 E_i}\left(\frac{\partial\mathrm{Re}\Sigma^{ret}_i}{\partial t} 
	+ \frac{E_i^2-\mathbf{P}_i^2-M_0^2-\mathrm{Re}\Sigma^{ret}_i}{\mathrm{Im}\Sigma^{ret}_i}
	\,\frac{\partial\mathrm{Im}\Sigma^{ret}_i}{\partial t}\right), \nonumber
\end{align}
where
\begin{equation}
	C_i = \frac{1}{2 E_i}\left(\frac{\partial\mathrm{Re}\Sigma^{ret}_i}{\partial E_i} 
	+ \frac{E_i^2-\mathbf{P}_i^2-M_0^2-\mathrm{Re}\Sigma^{ret}_i}{\mathrm{Im}\Sigma^{ret}_i}
	\,\frac{\partial\mathrm{Im}\Sigma^{ret}_i}{\partial E_i}\right).
\end{equation}
With the momentum independent self-energies described in Eq. (\ref{selfen}) the momentum derivatives of the self-energies can be dropped.
%\end{widetext} 

\subsection{Analytical prediction for homogeneous systems}
First let us analyze how the phase space coordinates of the individual particles and their mass distribution change, if we neglect the density fluctuations of the medium. If the density of the medium is constant in space, but changing in time, then the momenta of the particles are constant, and the masses of the particles change according to the following equation:
\begin{equation}
\label{masssquare}
	\frac{d}{dt}M_i(t)^2 =
	\left(\frac{d\mathrm{Re}\Sigma_i}{dt} 
	+ \frac{M_i^2(t)-M_0^2-\mathrm{Re}\Sigma_i}{\mathrm{Im}\Sigma_i}\,\frac{d\mathrm{Im}\Sigma_i}{dt}\right).
\end{equation}
Equation (\ref{masssquare}) can be reorganized to have the following form:
\begin{equation}
\label{masssquare1}
	\frac{d}{dt}(M_i(t)^2 - M_0^2-\mathrm{Re}\Sigma_i)=
	\frac{M_i^2(t)-M_0^2-\mathrm{Re}\Sigma_i}{\mathrm{Im}\Sigma_i}\,\frac{d\mathrm{Im}\Sigma_i}{dt}.
\end{equation}
It is worth making a few points here. According to Eq. (\ref{spectfunc}) the peak of the spectral function is at $M_0^2+\mathrm{Re}\Sigma(\mathbf{r},t)$.
If we have a testparticle with mass just at the peak of the distribution: $M_i(t)^2 = M_0^2+\mathrm{Re}\Sigma_i$, then it always stays at the peak of the spectral function independent of how the surrounding medium changes. We now see that the peak of the mass distribution during the whole evolution agrees with the peak of the spectral function calculated from the self-energies at the local densities.
The next question is what happens with the mass distribution itself during the 
evolution. After some algebraic manipulations of Eq. (\ref{masssquare1}) we get a relation, which can be integrated over time: 
\begin{equation}
\label{masssquare2}
	\frac{d}{dt}\log\left(M_i^2(t)-M_0^2-\mathrm{Re}\Sigma_i(t)\right) =
	\frac{d}{dt}\log\left( \mathrm{Im}\Sigma_i(t)\right).
\end{equation}
After integration we can express the particles masses at any time using the following relation \cite{Leupold-trans,Cassing3,Effenberger}:
\begin{equation}
\label{masssquare3}
	\frac{M_i^2(t)-M_0^2-\mathrm{Re}\Sigma_i(t)}{M_i^2(0)-M_0^2-\mathrm{Re}\Sigma_i(0)} =
	 \frac{\mathrm{Im}\Sigma_i(t)}{\mathrm{Im}\Sigma_i(0)}.
\end{equation}
Using this equation it is straightforward to prove that in the homogeneous case the mass distribution will remain a Breit-Wigner distribution, but with the self-energies changing according to the medium.  
\begin{align}
	f(M'^2,t)&=A(M^2,0,0)\frac{dM^2}{dM'^2} \nonumber \\
	&=A(M^2,0,0)\frac{\mathrm{Im}\Sigma(0,0)}{\mathrm{Im}\Sigma(0,t)}=A(M'^2,0,t)
\end{align}
Here we used the fact that at $t=0$ the spectral function $A(M^2,0,0)$ describes the distribution of the masses, and that the mass distribution function changes according to the change of the mass square intervals. The last step can be proven by using the explicit form of the Breit-Wigner distribution. Similar result was derived in \cite{Leupold-trans} after the gradient expansion for the imaginary part of the retarded propagator which plays the role of the mass distribution, however, up to our knowledge it was not deduced yet for the testparticle dynamics (Eq. (\ref{Eq-Hamilton})).

Our result tells us that if we simulate a mass distribution by testparticles according to the local spectral function, and evolve it using Eq. (\ref{Eq-Hamilton}), then the mass distribution will always be the Breit-Wigner distribution with the self-energies corresponding to the actual density. This means that in our model there is no memory effect at all concerning the mass distributions. When solving the Kadanoff-Baym equations themselves \cite{Schenke}, substantial memory effects were found. The equations of motion Eq. (\ref{Eq-Hamilton}) does not have this feature, so all the terms responsible for the memory effects present in the Kadanoff-Baym equations must have been neglected during the gradient expansion, as Eq. (3.33) in \cite{Leupold-trans} shows.

If we are interested only in the mass distribution of the resonances in a transport model, we can use the spectral function at the given density, we do not need testparticles and evolution equations for them to calculate the distribution. However, if the momentum distribution and velocities of the resonances also play role, then we should use Eq. (\ref{Eq-Hamilton}). Momentum distributions are important for example when one is interested in the produced dilepton spectrum. Also, when solving Eq. (\ref{Eq-Hamilton}) we gain a knowledge about how the vector mesons created in the dense matter get on-shell: we can find out how much of the final energy comes from the mesons initial kinetic energy, and how much energy comes from the medium.

\subsection{Change of mass distributions in a dynamical model}

Having seen in the previous subsection that in a homogeneous system the mass distribution of the vector mesons should always follow the spectral function of the medium without any delay, it is interesting to check whether it is true also in the simulations. In principle there can be differences due to density fluctuations. We test their influence numerically in the medium presented in the earlier chapters.

We initialize the medium from a $7\, \textrm{AGeV}$ ${}^{12}C+{}^{12}C$ collision, while constraining the particles to the $a=5\, \mathrm{fm}$ edge sized box. We let the medium equilibrate for $t_0 = 300\, \mathrm{fm/c}$, and then we put in $10000$ $\rho$ and $\omega$ testparticles with self-energies described in Eq. (\ref{selfen}), with random spatial coordinates, zero momentum, and masses chosen randomly according to the local Breit-Wigner distribution Eq. (\ref{spectfunc}). The density can be calculated from the number of baryons in the system and the box size, and turns out to be $\rho_0 = 0.192\, \textrm{fm}^{-3}$.

\begin{figure}[h] 
 \centering
   \includegraphics[width=\linewidth]{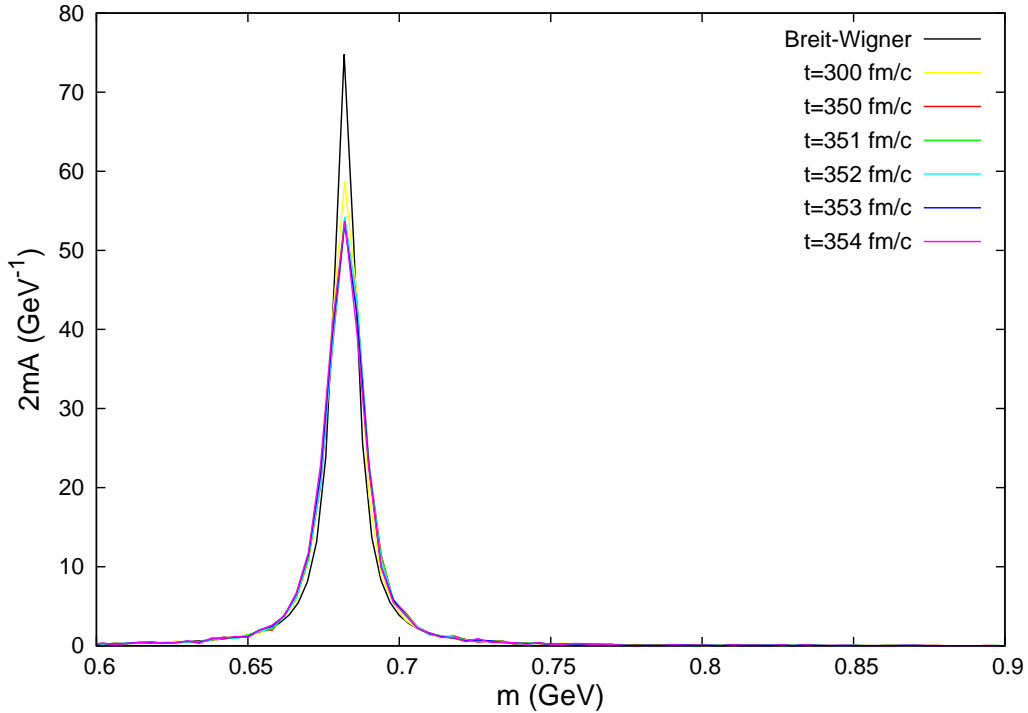}
   \caption{The distribution of masses of the $\omega$ mesons at different times compared to the Breit-Wigner distribution corresponding to the average density of the box. Parameters in the self-energy are chosen to be $M_x = 682\, \textrm{MeV}$, $\Gamma_x = 8.5\, \textrm{MeV}$.}
   \label{omegadist_spect_fluc}
\end{figure}
\begin{figure}[h] 
 \centering
   \includegraphics[width=\linewidth]{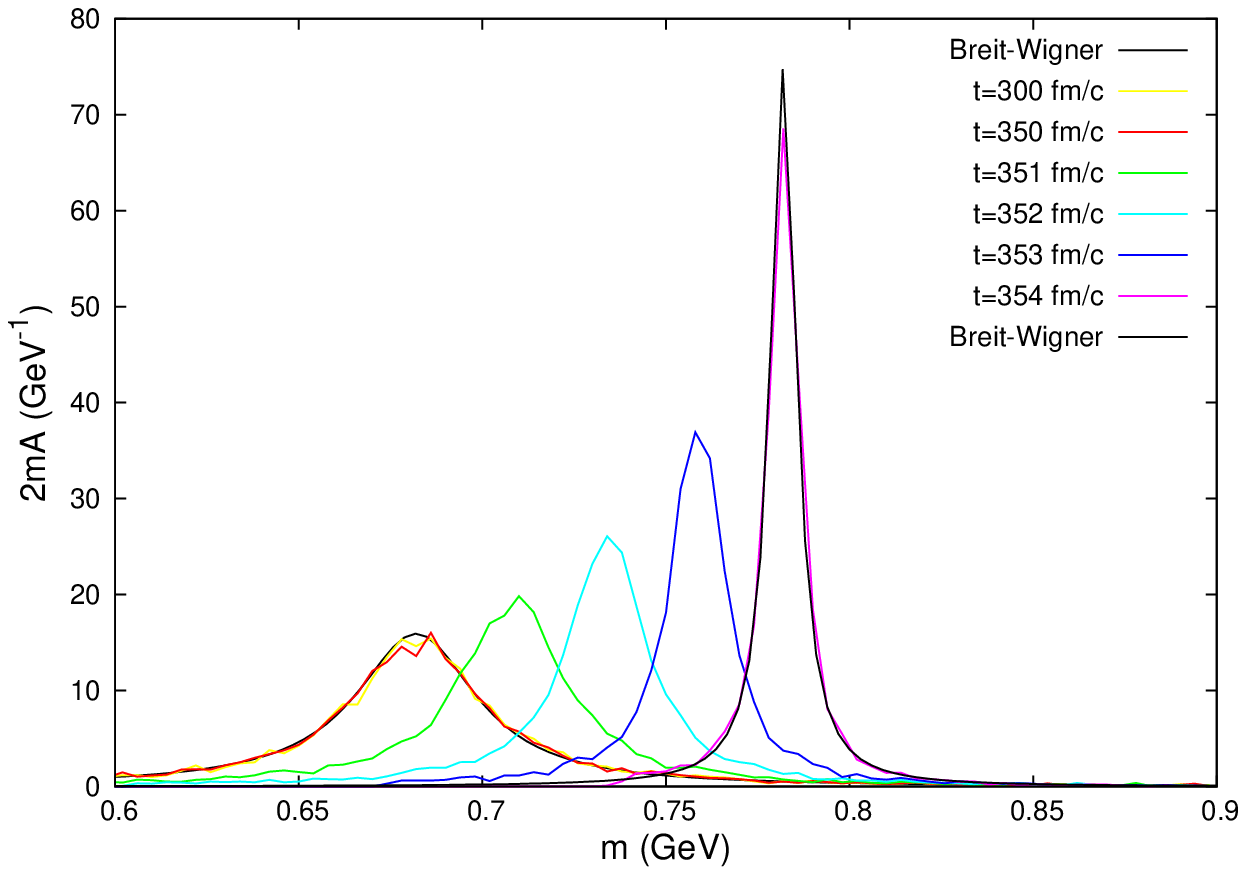}
   \caption{The distribution of masses of the $\omega$ mesons at different times compared to the Breit-Wigner distribution corresponding to the average density of the box. Parameters in the self-energy are chosen to be $M_x = 682\, \textrm{MeV}$, $\Gamma_x = 40\, \textrm{MeV}$. The density is decreased from $\rho_0$ to zero under $\Delta t=4\, \textrm{fm/c}$.}
   \label{omegadist_spect_chall}
\end{figure}
\begin{figure}[h]
 \centering
   \includegraphics[width=\linewidth]{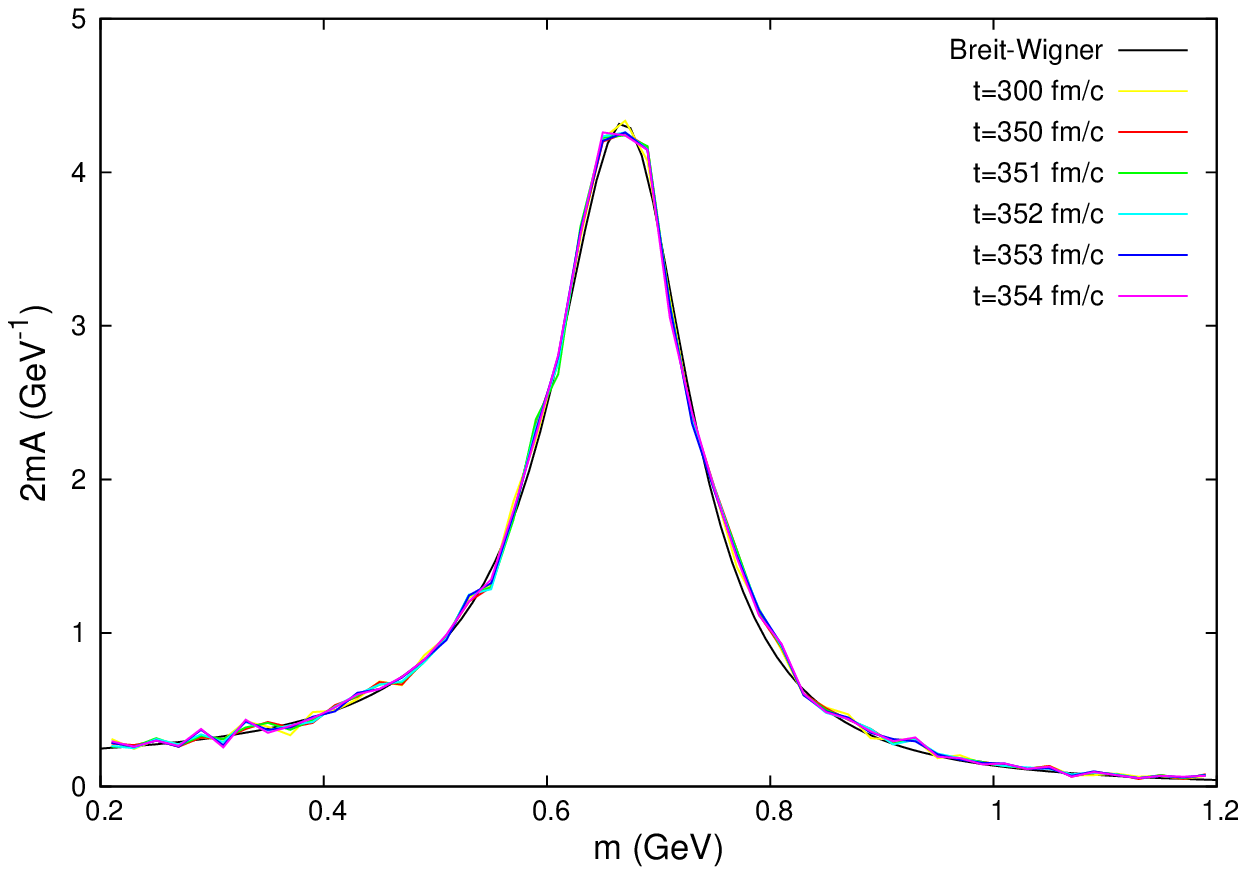}
   \caption{The distribution of masses of the $\rho$ mesons at different times compared to the Breit-Wigner distribution corresponding to the average density of the box. Parameters in the self-energy are chosen to be $M_x = 668\, \textrm{MeV}$, $\Gamma_x = 150\, \textrm{MeV}$.}
   \label{rhodist_spect_fluc}
\end{figure}
\begin{figure}[h]
 \centering
   \includegraphics[width=\linewidth]{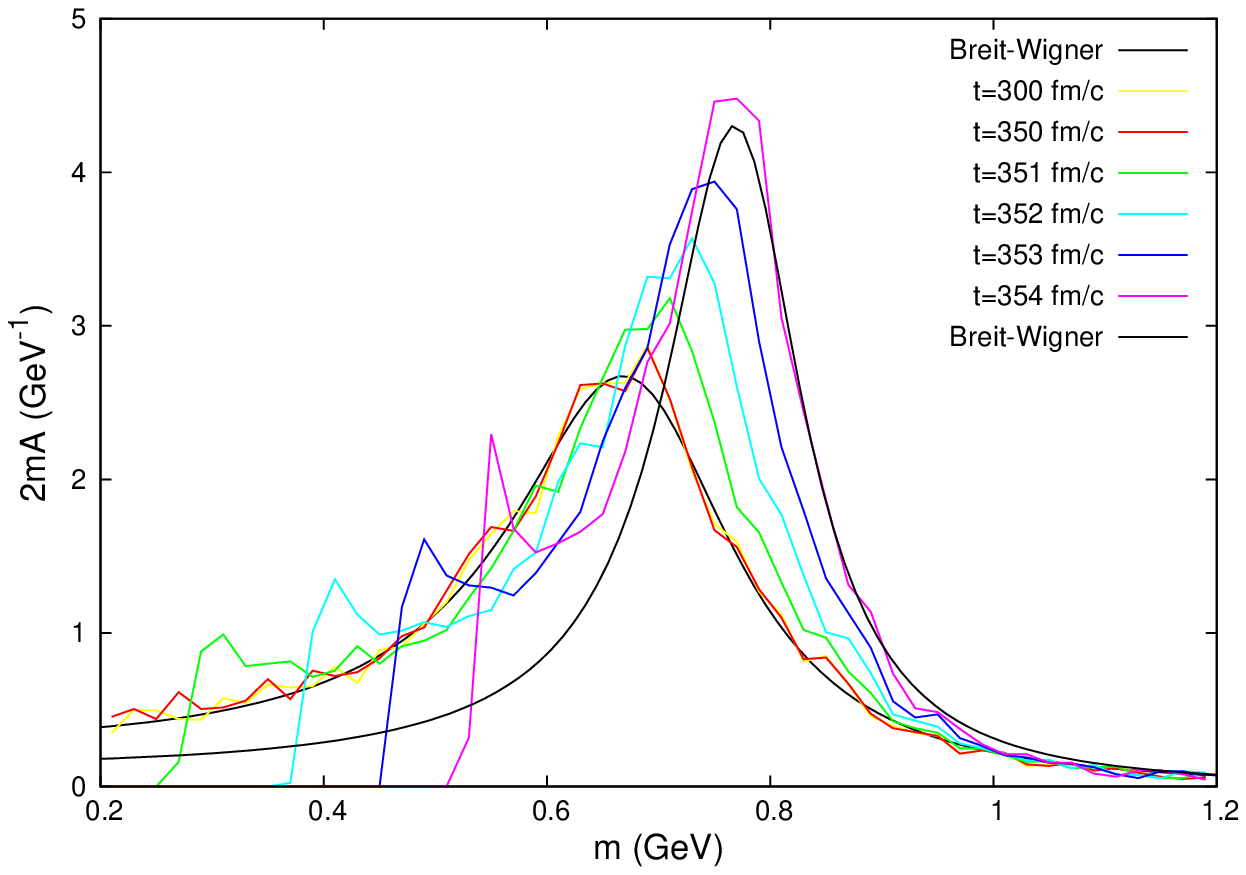}
   \caption{The distribution of masses of the $\rho$ mesons at different times compared to the Breit-Wigner distribution corresponding to the average density of the box. Parameters in the self-energy are chosen to be $M_x = 668\, \textrm{MeV}$, $\Gamma_x = 250\, \textrm{MeV}$. The density is decreased from $\rho_0$ to zero under $\Delta t=4\, \textrm{fm/c}$.}
   \label{rhodist_spect_chall}
\end{figure}

At first, we did not change the density of the medium, and we checked if the mass distribution changes. We chose the self-energy parameters to be $M_x^{\omega} = 682\, \textrm{MeV}$, $\Gamma_x^{\omega} = 8.5\, \textrm{MeV}$, $M_x^{\rho} = 668\, \textrm{MeV}$ and $\Gamma_x^{\rho} = 150\, \textrm{MeV}$, so we shifted their peak from the vacuum value by $100\, \textrm{MeV}$ and we did not change their width. We can see in Fig. \ref{omegadist_spect_fluc} that in the case of $\omega$ mesons even the initial mass distribution differs significantly from the Breit-Wigner distributions. This is due to density fluctuations: the average of different density Breit-Wigner distributions does not give back the Breit-Wigner distribution of the average density, but we get a broader distribution. Then the mesons gain some momentum (they were initialized with zero momentum), and the distribution broadens a little even more. For $\rho$ mesons in Fig. \ref{rhodist_spect_fluc} this effect cannot be seen: the mass distribution agrees well with the Breit-Wigner distribution. Hence we can draw the conclusion, that density fluctuations are only important in the case of very narrow resonances.

%\begin{figure}[h] 
% \centering
%   \includegraphics[width=\linewidth]{omegadist_spect_chall.eps}
%   \caption{The distribution of masses of the $\omega$ mesons at different times compared to the Breit-Wigner distribution corresponding to the average density of the box. Parameters in the self-energy are chosen to be $M_x = 682\, \textrm{MeV}$, $\Gamma_x = 40\, \textrm{MeV}$. The density is decreased from $\rho_0$ to zero under $\Delta t=4\, \textrm{fm/c}$.}
%   \label{omegadist_spect_chall}
%\end{figure}
%\begin{figure}[h]
% \centering
%   \includegraphics[width=\linewidth]{rhodist_spect_chall.eps}
%   \caption{The distribution of masses of the $\rho$ mesons at different times compared to the Breit-Wigner distribution corresponding to the average density of the box. Parameters in the self-energy are chosen to be $M_x = 668\, \textrm{MeV}$, $\Gamma_x = 250\, \textrm{MeV}$. The density is decreased from $\rho_0$ to zero under $\Delta t=4\, \textrm{fm/c}$.}
%   \label{rhodist_spect_chall}
%\end{figure}

Now we discuss the evolution of mass distributions in the case of changing 
density. In our model of strongly interacting matter this can be achieved by 
the removal of baryons from the system. We decreased the density from $\rho_0$ 
to zero under $\Delta t=4 \textrm{fm/c}$ linearly. In this simulation we chose 
the self-energy parameters to be $M_x^{\omega} = 682\, MeV$, 
$\Gamma_x^{\omega} = 40\, MeV$, $M_x^{\rho} = 668\, MeV$ and 
$\Gamma_x^{\rho} = 250\, MeV$, so we shifted their peak from the vacuum value by 
$100\, \textrm{MeV}$ and we also modified their widths relative to their vacuum 
value. In Fig. \ref{omegadist_spect_chall} and \ref{rhodist_spect_chall} we can 
see, that at the creation time both the $\omega$ and $\rho$ mass distribution 
agrees well with the Breit-Wigner formula, and in the case of the $\omega$ 
mesons, this will also be the case after the removal of the nuclear medium. 
This is in agreement with our analytical considerations in the previous 
subsection. However in the case of the $\rho$ mesons the mass distribution 
after the removal of the nuclear medium differs significantly from the 
Breit-Wigner distribution. The peak is well described by the Breit-Wigner 
formula, however, the low mass tail differs. The reason for this difference is 
the following: a testparticle with a very small mass in vacuum would have 
negative $M^2$ in the medium according to Eq. (\ref{masssquare3}), and in our 
model we do not simulate such particles. The Breit-Wigner formula cannot be 
exactly implemented, since that would require negative $M^2$ testparticles. The 
$M^2>0$ initial cutoff will result in higher $M^2$ cutoffs in the more and more 
dilute medium, since the distribution gets narrower. The net effect of the difference on the low mass tail is rather small for physical observables like dilepton production since the time average of the simulated distribution function and the time average of the exact Breit-Wigner distributions do not differ substantially.

Our simulations thus suggest, that for resonances whose in-medium decay width 
is not very small (larger than $20\, \textrm{MeV}$), the mass distribution of 
the particles can always be well described by the spectral function 
corresponding to the actual density of the system. Differences can occur at the 
low mass tail of the distribution, but the peak is well described by the 
Breit-Wigner distributions even for broad resonances. This suggests that the 
mass distribution of vector mesons always follows the spectral function of the 
nuclear matter, just as it was suggested from the analysis describing 
homogeneous nuclear matter.

\section{Summary \label{Summary}}

In summary we have considered the equilibration in a relativistic heavy ion collision using our transport model. We applied periodic boundary conditions to close the system in a box. We found that the thermal equilibration takes place in the first 20-40 fm/c which time is comparable to the duration of a heavy ion collision. The chemical  equilibration is a slower process. The nonstrange degrees of freedom approximately equlibrate in 50-100 fm/c, but for the strangeness the equilibration time is from 100 to 300 fm/c depending on the density and on the bombarding energy.

We also study the propagation of broad resonances
within our approach. Vector mesons are described by spectral functions
and these are evolved in space and time by a testparticle method.
We have utilized here the transport equations from Ref.~\cite{Cassing1,Leupold-trans}
which are approximations of the much more involved Kadanoff-Baym equations \cite{Schenke}. We solved the equations analytically in the case of a homogeneous system. We found that in our model the mass distributions of the resonances always follow the spectral function corresponding to the given density, so there is no memory effect in contrast to the exact treatment. The same results were obtained by the numerical simulations as well. This means that the mass distributions of resonances can be extracted without following a large number of testparticles. Our results however do not mean that Eqs. (\ref{Eq-Hamilton}) are useless: they describe the momentum evolution of the resonances, which is important for example for calculating dilepton production since the time spent in the dense phase depend on its momentum.

We acknowledge stimulating discussion with Stefan Leupold, Carsten Greiner and Volker Metag. This work were supported by the Hungarian OTKA funds NK101438 and K109462 and  by the HIC for FAIR within the framework of the LOEWE program launched by the State of Hesse.

%\References

%\begin{thebibliography}
\bibliographystyle{elsarticle-num} 
\bibliography{refs}
%\end{thebibliography}

\end{document}